\documentstyle[twocolumn,aps,prl]{revtex}
\begin{document}
%
%
\wideabs{
%
%
\title{Simple model of a limit order-driven market.}
\author{Sergei Maslov}
\address{Department of Physics, Brookhaven National Laboratory,
Upton, NY 11973}

\date{\today}
\maketitle
%
%
\begin{abstract}
We introduce and study a simple model of a 
limit order-driven market. Traders in this model can either 
trade at the market price or place a limit order, i.e. an 
instruction to buy (sell) a certain amount of the stock if its 
price falls below (raises above) a predefined level.
The choice between these two options is purely random (there are no
strategies involved), and the execution price of a limit order 
is determined simply by offsetting the most recent market 
price by a random amount. Numerical simulations of this model 
revealed that despite such minimalistic rules the price pattern
generated by the model has such realistic features as ``fat'' tails of
the price fluctuations distribution,
characterized by a crossover between two power law exponents, 
long range correlations of the volatility, and a non-trivial Hurst
exponent of the price signal. 
\end{abstract}
%
%
}
\narrowtext

Recent years have witnessed an explosion of activity in the area of 
statistical analysis of high-frequency 
financial time series \cite{books}. 
This lead to the discovery of 
robust and to a certain degree universal features of 
price fluctuations, and triggered theoretical studies 
aimed at explaining or simply mimicking these observations. 
The list of empirical facts that need to be
addressed by any successful theory or model is: 

(i) The histogram of short time-lag increments of market price 
has a very peculiar non-Gaussian shape with a sharp maximum and broad 
wings \cite{Mandelbrot}. The current consensus about the functional form of this 
distribution is that up to a certain point it
follows a Pareto-Levy distribution, with the exponent of its 
power law tail $1+\alpha_1 \sim 2.4-2.7$, 
after which it crosses over to a steeper power law  
$1+\alpha_2 \sim 4-4.5$ \cite{Gopikrishnan_Stanley}, 
or, as reported in another study \cite{Cont_Bouchaud,Mantegna_Stanley}, 
to an exponential decay.

(ii) When viewed on time scales less than several trading days, 
the graph of price vs time appears to have a  
Hurst exponent $H \simeq 0.6 -0.7$ \cite{Mandelbrot,Mantegna_Stanley}, 
different from an ordinary uncorrelated random walk value $H_{RW}=0.5$. 

(iii) The volatility
(the second moment of price fluctuations) exhibits correlated
behavior. It is manifested in clustering 
of volatility, i.e. the presence of regions of unusually high
amplitude of fluctuations
separated by relatively quiet intervals, visible with a ``naked eye'' 
in the graph of price increment vs time. These  
clustering effects determine the shape of the autocorrelation 
function of volatility as a function of time, which was shown to 
decay as a power law with a very small exponent 
$\gamma\simeq 0.3-0.4$ and no apparent cutoff 
\cite{Cont_Bouchaud,Liu_Stanley}.

There are several approaches to modeling market mechanics.
In one type of models price fluctuations 
result from trading activity of conscious
agents, whose decisions to buy or sell are dictated by well defined 
strategies.  These strategies evolve in time (often according to some 
Darwinian rules) and give
rise to a slowly changing fluctuation pattern. 
There is little doubt that the evolution and dynamics of investor's 
strategies and beliefs influence the long term behavior of real market prices. 
For example, if some company could not keep up with the competition, 
sooner or later investors would realize it, and in the  
long-term its stock price would go down. 
However, it is unclear how does it influence the properties of stock  
price fluctuations at very short timescales, which do not allow 
time for traders to update their strategies or for a company to change
its profile. Another problem with models explaining short time 
price fluctuations in terms of strategy evolution  
is that they inevitably lead their creators to shaky grounds of
speculations about relevant and irrelevant psychological
motivations of a ``typical'' trader in a highly 
heterogeneous trader population. The remarkable universality 
of general features of price fluctuations in markets of 
different types of risky assets
such as stocks, options, foreign currency, and
commodities (say, cotton or oil) makes one to suspect
that in fact psychological factors play little role in determining 
their short time properties, and leads one to try to look for
a simpler mechanisms giving rise to these features.

In this work we do a first step in this direction by introducing
and numerically studying a simple market model, where 
a nontrivial price pattern arises not 
due to the evolution of trading strategies, 
but rather as a consequence of {\it trading rules themselves}.
Before we proceed with formulating the rules of our model we need to
define several common market terms.
A market trader is usually allowed to place 
a so-called ``limit order to sell (buy)'', which is an instruction to 
automatically sell
(buy) a particular amount of stock if its market price would 
raise higher (or drop lower for a limit buy order) than the
predetermined threshold. This threshold is sometimes referred to as
the execution price of the limit order. In many modern markets, 
known to economists as order-driven markets \cite{yang_pdf}, 
limit orders placed by ordinary traders
constitute the major source of the liquidity of the market.  
It means that a request to {\it immediately} buy or sell a particular amount
of stock at the best market price, or ``market order'',  
is filled by matching it to 
the best unsatisfied limit order in the limit order book.
To better understand how transactions are made at an order-driven market it
is better to consider the following simple example: suppose one trader
(trader \#1) has submitted a limit order to sell 1000 shares of the stock of 
a company X, provided that its price would exceed \$20/share. Subsequently another
trader (trader \#2) has submitted a limit order to sell 2000 shares of X 
if the price would exceed \$21/share. Finally, a third trader decides
to buy 2000 shares of X at the market price. In the absence
of other limit orders his order will be filled as follows:
he will buy 1000 shares from trader \#1 at \$20/share 
and 1000 shares from trader \#2 at
\$21/share. After this transaction the limit order book would contain
only one partially filled limit order to sell, that of trader \#2 
to sell 1000 shares of X at \$21/share. 

Traders in our model can either trade stock {\it at the market 
price} or place a {\it limit order to sell
or buy}. To simplify the rules of our toy market,  
traders are allowed to trade only one unit (lot) of stock in each 
transaction. That makes all limit and market orders to be of 
the same size. The empirical study of limit-order trading at
the ASX \cite{yang_pdf} can be used to partially justify this 
simplification. In this work it was observed that limit orders mostly come in 
sizes given by round numbers such as 1000 shares and (to a lesser extent) 
10000 shares.
Unlike many other market models, we do not fix the number 
of traders. Instead, at each time step a ``new'' trader
appears out of a ``trader pool'' and attempts to make a transaction.
{\it With equal probabilities this new trader is a seller or a buyer.} 
He then performs one of the following two actions: 
\begin{itemize}
\item with probability $q_{ lo}$ he places a
limit order to sell/buy.
\item otherwise (with probability $1-q_{ lo}$) he trades (sells or 
buys) at the market price. 
\end{itemize}
The rule of execution of a market order in our model is particularly
simple. Since all orders are of the same size, a market
order is simply filled with the best 
limit order (i.e. the highest bid among limit orders to buy and the
lowest ask among limit orders to sell), which is subsequently 
removed from the limit order book. 
This transaction performed at the execution price of the
best limit order sets a new value of the market price $p(t)$.

To complete the definition of the rules one needs to specify
how a trader who selected to place a new limit order decides on its
execution price. Traders in our model do this in a very
``non-strategic'' way by simply offsetting 
the price of the last transaction performed on the market 
(current market price $p(t)$), by a random number $\Delta$.
This {\it positive} random number is drawn each time from the same 
probability distribution $P(\Delta)$. A new limit order to sell 
is placed {\it above the current price} at $p(t)+\Delta$, while 
a new limit order to buy -- below it at $p(t)-\Delta$. 
This way ranges of limit orders to sell and to buy never overlap, 
i.e. there is always a
positive gap between highest bid and lowest ask prices.
This ``random offset'' rule constitutes a reasonable 
first order approximation to what may happen in real order-driven 
markets and is open to modifications if it fail a reality check. 
The most obvious variants of this rule, which we plan to
study in the near future, are i) A model 
where each trader has his individual distribution $P(\Delta)$. This
modification would allow for the coexistence of ``patient'' traders who
do not care very much about when their order will be executed or if it
will be executed at all, and can therefore select large $\Delta$ and
pocket the difference, and ``impatient'' traders who need 
their order to be executed soon, so they tend to select a small 
$\Delta$ or trade at the market price.
ii) A model in which the probability distribution
 of $\Delta$ is determined by the historic 
volatility of the market. This rule seems to be particularly
reasonable description of a real order-driven market. 
Indeed, if traders  selection of $\Delta$ is influenced primarily by
his desire to reduce waiting time before his order is executed, 
then it would make sense to select a 
larger $\Delta$ in a more volatile market, which is likely to 
cover larger price interval during the same time interval. 
However, before any of these more complicated versions of this rule
could be explored one needs to study and understand the behavior of
the base model, where $\Delta$ is just a random number, uncorrelated
with volatility and/or the individual trader profile. 

One should notice that the behavior of traders in our model is
completely passive and ``mechanical'': once a limit order is placed 
it cannot be removed or shifted in response to a current market 
situation. This makes our rules fundamentally different from 
these of the Bak-Paczuski-Shubik (BPS) model \cite{Bak97}, where 
randomly increases or decreases their quotes at each time step. 
Such haphazard trader behavior cannot be realized in 
an order-driven market, where each change of the limit-order 
execution price carries a fee.

We have simulated our model with $q_{ lo}=1/2$, 
i.e. when on average half of the traders select to place limit orders,  
while the other half trade at the market price. 
The random number $\Delta$, used in setting an execution price of a new
limit order,  was drawn from a uniform distribution in the interval 
$0 \leq \Delta \leq \Delta_{\rm max}=4$. Obviously, price patterns in models 
with different values of $\Delta_{\rm max}$ are identical up to an
overall rescaling factor. Our choice of $\Delta_{\rm max}=4$ was 
dictated by the desire to compare the behavior of the model 
with continuous spectrum of $\Delta$ to that with a discrete 
spectrum $\Delta=\{1,2,3,4\}$. Discrete spectrum of $\Delta$ 
may better compare to the behavior of  real markets, 
where all prices are multiples of
a unit tick size. Our comparison confirmed that most scaling
properties of the price pattern are the same in both variants.
We were surprised to notice that non-trivial features of our 
model survived even in a model with deterministic $\Delta=1$. 
To improve the speed of numerical simulations we studied a 
version of the model, where only $2^{17}$
lowest ask and highest bid quotes were retained. The list of quotes was
kept ordered at all times, which accelerated the search for the highest
bid and lowest ask limit orders whenever a transaction at the market price was 
requested. We also studied a variant of our model where each limit
order had an expiration time: if a limit order was not filled 
within 1000 time steps it was removed from the list. 
Not only this rule prevented an occasional 
accumulation of a very long list of limit orders, but also 
it made sense in terms of how limit orders are organized in a real market. 
Indeed, limit orders at, for example, New York Stock Exchange 
are usually valid only during the trading day when
they were submitted. There are also so called ``good till canceled'' (or
open) orders, which are valid until they are executed or withdrawn 
\cite{nyse_www}. 
Then the version of our model, where the expiration time of a limit
order is not specified  
corresponds to all orders being ``good till canceled'', while
the version, where only the most recent orders are kept,
mimics the market composed of only ``day orders''. 
We have checked
that for any reasonably large value of the cutoff parameter, no matter if
it is an expiration time or the number of best sell/buy orders to 
keep, one ends up with the same scaling properties 
of price fluctuations.    

In Fig. 1 we present an example of price history in 
one of the runs of our model. Visually it is clear that this 
graph is quite different from an ordinary random walk.
This impression is confirmed by looking at the pattern of price increments
$p(t+1)-p(t)$, shown in the same figure. One can see that 
large increments are clustered in regions of high volatility,
separated by relatively quite intervals.
The Fourier spectrum of the price signal averaged over many runs of the 
models provides us with a value of the Hurst
exponent $H$ of the price graph. Indeed, the exponent of the 
Fourier transform of price autocorrelation function $S_p(f)$ is 
related to the Hurst exponent as $S_p(f) \sim f^{-(1+2H)}$. 
The log-log plot of $S_p(f)$, logarithmically binned and 
averaged over multiple realizations of the 
price signal of length $2^18$, is shown in the
inset to Fig. 1. It has an exceptionally clean $f^{-3/2}$
functional form for over 5 decades in $f$, which corresponds to the Hurst
exponent of the price signal $H=1/4$. This exponent 
is definitely different from its random walk
value $H_{RW}=1/2$. A Hurst exponent $H=1/4$
was also observed in the Bak-Paczuski-Shubik model A \cite{Bak97,Tang99}.
An intuitive argument in favor of a small Hurst exponent can be constructed 
for our model.  According to the rules of the model 
an execution price of a new 
limit order is always determined relative to the current price. 
It is also clear that a large density of limit orders around 
current price position reduces its mobility. 
Indeed, in order for the price to move to the new position all
limit orders in the interval between the current and new values of the
price must be filled by market orders. 
If for one reason or the other the price remained
fairly constant for a prolonged period of time, limit orders created
during this time tend to further trap the price in this region. 
This self-reinforcing mechanism qualitatively explains the 
slow rate of price change in our model. 
Unfortunately, the nontrivial Hurst exponent $H=1/4$ 
is a step in the wrong direction from its random walk
value $H_{RW}=1/2$. Indeed, the short time Hurst
exponent of real stock prices was measured to be 
$H_{real} \simeq 0.6-0.7$.

The amplitude of price fluctuations in our model has significant long
range correlations. One natural measure of these correlations is
the autocorrelation function of the {\it absolute value} of 
price increments $S_{abs}(t)=\langle 
|p(t'+t+1)-p(t'+t)||p(t'+1)-p(t')|\rangle_{t'}$\cite{Liu_Stanley}. 
In our model this quantity was measured to have a power law tail 
$S_{abs}(t) \propto t^{-1/2}$. This is 
illustrated in Fig. 2 where the Fourier transform of $S_{abs}(t)$ 
has a clear $f^{-1/2}$ form. The 
exponent $\gamma=1/2$ of $S_{abs}(t)  \propto t^{-\gamma}$ in our model 
is not far from  $\gamma=0.3$ measured in the S\&P 500 stock index 
\cite{Liu_Stanley}. In Fig. 2 we also show the Fourier transform of the 
autocorrelation function of {\it signs} of price increments 
$S_{sign}(t)=\langle 
sign[p(t'+t+1)-p(t'+t)]sign[p(t'+1)-p(t')]\rangle_{t'}$, which has a 
white noise (frequency independent) form. This is again, similar 
to the situation in real market, where signs of price
increments are known to have only short range ($< 30 $ min) 
correlations. 

Finally, in Fig. 3 we present three histograms of price increments
$p(t+\delta t)-p(t)$ in our
model, measured with time lags $\delta t=1,10,100$. The overall form of
these histograms is strongly non-Gaussian and is reminiscent of the 
shape of such distribution for real stock prices.  As the time
lag is increased the sharp maximum of the distribution 
gradually softens, while its wings remain strongly non-Gaussian.
In the inset we show a log-log plot of the histogram of 
$p(t+1)-p(t)$ ($\delta t=1$) collected during $t_{stat}=3.5 \times 10^7$ 
timesteps (as compared to $t_{stat}=40000$ for the data shown 
in the main panel) and logarithmically binned. 
One can clearly distinguish two power law regions separated by a
sharp crossover around $p(t+1)-p(t) \simeq 1$. The exponents of 
these two regions were measured to be $1+\alpha_1 =0.6 \pm 0.1$ and 
$1+\alpha=3 \pm 0.2$. The power law exponent $1+\alpha_2=3$ of the 
far tail lies right at the borderline, separating the Pareto-Levy region
$\alpha<2$, where the distribution has an infinite second moment,  
from the Gaussian region. In any case, since price fluctuations 
in our model were shown
to have long range correlations, one should not expect convergence of
the price fluctuations distribution to a universal Pareto-Levy or Gaussian 
functional form as $\delta t$ is increased.  The existence of a similar
power law to power law crossover was reported in the distribution
of stock price increments in NYSE, albeit with different exponents 
$1+\alpha_1 \simeq 1.4-1.7$, and $1+\alpha_2 \simeq 4-4.5$ 
\cite{Gopikrishnan_Stanley}. The mechanism 
responsible for this crossover in a real market is at present
unclear. 

In conclusion, we have introduced and numerically studied a simple
model of a limit order-driven market, where agents randomly 
submit limit or market orders. The execution price of new 
limit orders is set by offsetting the current market 
price by a random amount. In spite of such strategy-less, 
mechanistic behavior of traders, the price time series in 
our model exhibit a highly nontrivial behavior 
characterized by long range correlations, fat tails in the histogram of 
its increments, and a non-trivial Hurst exponent. 
These results are in qualitative agreement 
with empirically observed behavior of prices on real stock
markets. More work is required to try to modify the rules of
our model in order to make this agreement more quantitative.

The work at Brookhaven National Laboratory was 
supported by the U.S. Department of Energy Division
of Material Science, under contract DE-AC02-98CH10886.
The author thanks Y.-C. Zhang for useful discussions, and 
the Institut de Physique Th\'eorique,
Universit\'e de Fribourg for the hospitality and financial support 
during the visit, when this work was started.

\begin{figure}
\caption{The price signal $p(t)$ and its derivative $\delta
p=p(t+1)-p(t)$ as a function of time. The inset shows the 
Fourier transform of the autocorrelation function of 
the price signal. Solid line is a fit 
$S_p(f)=f^{-(1+2H)}=f^{-3/2}$, corresponding to the Hurst exponent 
$H=1/4$.}
\end{figure}

\begin{figure}
\caption{Fourier transforms of autocorrelation functions of 
signs of price increments ($\times$) and absolute values of price
increments ($\bullet$) averaged over 700 realization of price record $2^{18}
\sim 2.6 \times 10^5$ time steps long.}
\end{figure}

\begin{figure}
\caption{Histograms of price increments $p (t)=
p(t+\delta t)-p(t)$ with time-lags $\delta t=1$ ($\circ$), 
10 ($\bullet$), and  100 ($\times$). The 
inset shows the histogram of positive price increments 
$p(t+1)-p(t)>0$ (negative increments have a virtually
indistinguishable hystogram) on a log-log plot. Power law 
fits in two regions give exponents $1+\alpha_1 =0.6$ and 
$1+\alpha=3$.}
\end{figure}

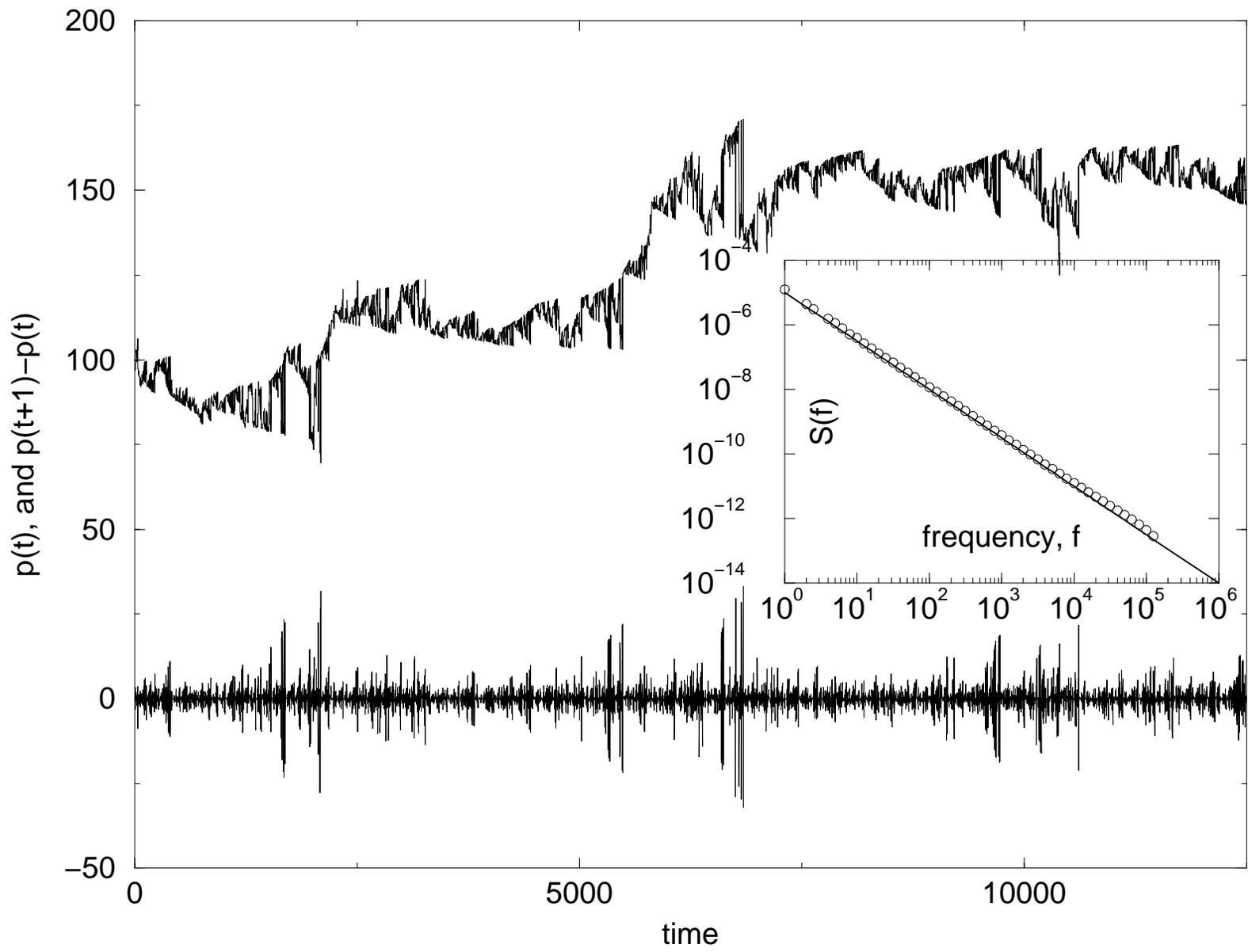
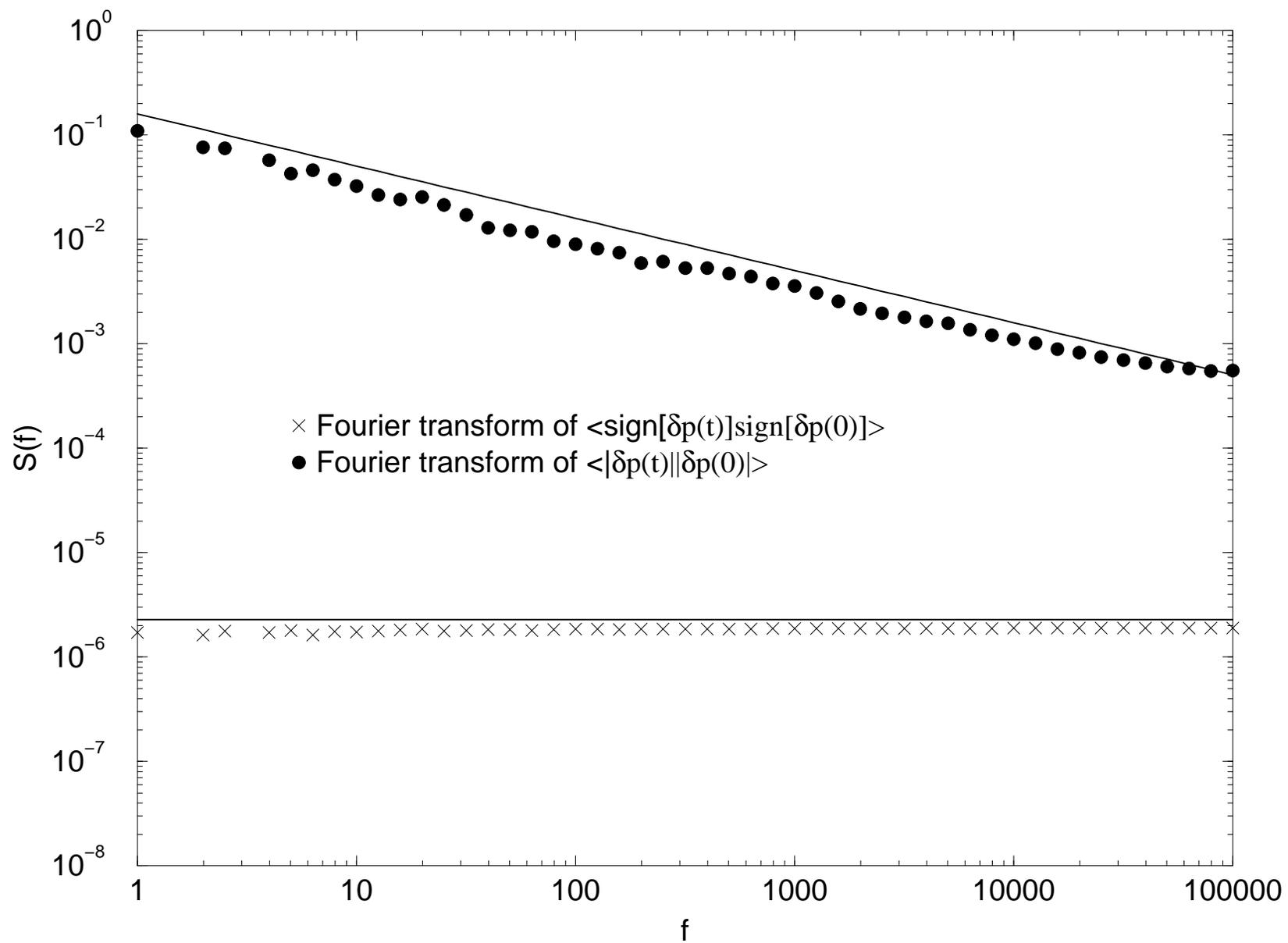
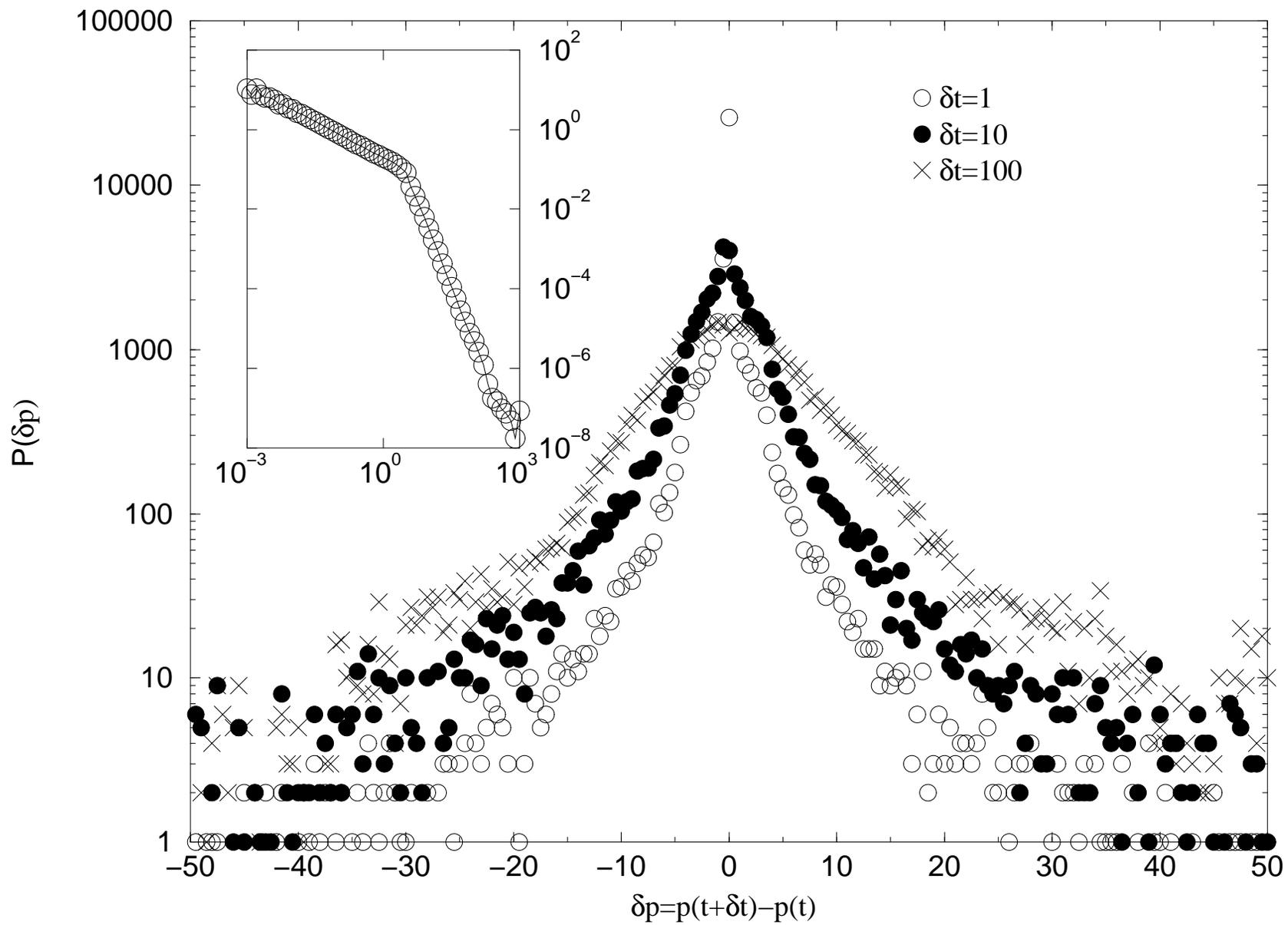
\end{document}